\documentclass[pre,groupedaddress,amsmath,amssymb,superscriptaddress,showpacs,twocolumn]{revtex4-1} 
\usepackage{natbib}
\usepackage{dcolumn}
\usepackage[pdftex]{graphicx}
\usepackage{epstopdf}
\usepackage{subfigure}
\usepackage{bm}
\usepackage[colorinlistoftodos,disable]{todonotes} 
\usepackage[pdftex]{hyperref}
\usepackage{soul}
\usepackage{nomencl} 

\newcounter{todoListItems}

\def\mys#1{{\mbox{\scriptsize{#1}}}}    

\newcommand{\degc}{\ensuremath{^{\circ}}\mathrm{C}}
\def\rg{r_{\mys{g}}}            
\def\a{a}                       
\def\cp{c_{\rm{p}}}             
\def\cpstar{c_{\rm{p}} / c_{\rm{p}}^{*} }             
\def\conc{\; \rm{mg}/\rm{ml}}   
\def\phic{\phicolloid}          
\def\phicolloid{\phi}           
\def\qR{\rg / \a}               
\def\Uc{\U_{\mys{c}}/\kb T}     
\def\U{U}                       
\def\D{\delta/ \a}              
\def\etaL{\eta_{\mys{L}}}       
\def\vis0{\eta_{0}}             
\def\R{R_{\mys{c}}}             
\def\q1{\left <q \right >}      
\def\h0{h_{0}}                  
\def\d0{d}                      
\def\hf{h_{\mys{f}}}            
\def\tauc{\tau_{\mys{c}}}       
\def\taud{\tau_{\mys{d}}}       
\def\tw{t_{\mys{w}}}            
\def\ts{\tau}                   
\def\tauesc{\tau_{\mys{esc}}}   
\def\vc{\nu}                    
\def\betac{\beta}               
\def\sig{\sigma}                
\def\sigg{\sigma_{\mys{g}}}     
\def\sigy{\sigma_{\mys{y}}}     
\def\lg{l_{\mys{g}}}            
\def\phig{\phicolloid_{\mys{g}}}
\def\alp{\alpha}                
\def\nbreak{N_{\mys{break}}}    
\def\nlink{N_{\mys{link}}}      
\def\elastic{K}                 

\def\kb{k_{\mys{B}}}            
\def\kBT{\kb T}                 
\def\Diff{D_{\mys{0}}}          
\def\Dt{D_{\mys{t}}}            
\def\Ds{D_{\mys{g}}}            

\makenomenclature 

\begin{document}

\bibliographystyle{apsrev}

\title{Sudden  collapse of a colloidal gel}

\author{Paul \surname{Bartlett}}
\email[Corresponding author:]{P.Bartlett@bristol.ac.uk}
\affiliation{School of Chemistry, University of Bristol, Bristol
BS8 1TS, UK.}
\author{Lisa J. \surname{Teece}}
\affiliation{School of Chemistry, University of Bristol, Bristol
BS8 1TS, UK.}
\author{Malcolm A. \surname{Faers}}
\affiliation{Bayer CropScience AG, 40789, Monheim am Rhein, Germany}


\begin{abstract}

Metastable gels formed by weakly attractive colloidal particles display a distinctive two-stage time-dependent settling behavior under their own weight. Initially a space-spanning network is formed that for a characteristic time, which we define as the lag time $\taud$, resists compaction. This solid-like behavior persists only for a limited time. Gels whose age $\tw$ is greater than $\taud$ yield and suddenly collapse. We use a combination of confocal microscopy, rheology and time-lapse video imaging to investigate both the process of sudden collapse and its microscopic origin in an refractive-index matched emulsion-polymer system.  We show that the height $h$ of the gel in the early stages of collapse is well described by the surprisingly simple expression, $h(\ts) = \h0 - A \ts^{\frac{3}{2}}$, with $\h0$ the initial height and $\ts = \tw-\taud$ the time counted from the instant where the gel first yields. We propose that this unexpected result arises because the colloidal network progressively builds up internal stress as a consequence of localized rearrangement events which leads ultimately to collapse as thermal equilibrium is re-established.

\end{abstract}

\maketitle
 \makeatletter
    \providecommand\@dotsep{5}
  \makeatother
  \listoftodos\relax

\section{Introduction}
\label{sec:introduction}

Soft materials such as colloidal suspensions and emulsions form a remarkably rich variety of non-ergodic states \cite{4213,7444,11922} -- examples of which are familiar to us in our daily life in products as diverse as  foodstuffs, surface coatings, fabric conditioners, and pesticides.  Out-of-equilibrium phases occur when suspensions are quenched deep into a region of thermodynamic phase separation. Unable to phase separate, amorphous solids form which are mechanically rigid  but without the long-range translational order characteristic of crystalline solids. Slow relaxation dynamics prevents the system  from reaching their underlying global equilibrium configurations so these amorphous solids evolve slowly in a complex energy landscape with a high number of local minima and as a result display glassy dynamics with a rich phenomenology of effects such as aging, non-linear responses, and spatial and temporal dynamic heterogeneities.

One of the most dramatic macroscopic manifestations of aging is the phenomenon of sudden network collapse in gels. Gels consist of a  network of particles linked together by long-lived attractive bonds. Sedimentation or creaming of the particles within a gel imposes a buoyant stress on the network which since gels are intrinsically rather delicate has dramatic consequences for microscopic structure and dynamics \cite{12069,8892,5711}. Weak gels, where the strength of the attractive potential at contact $\U_{\mys{c}}$ is only a few $\kBT$, show, for instance, a very unusual  mechanical response. Initially, the gel behaves as a solid but after a finite lag time $\taud$, the gel yields and catastrophically collapses. Sudden or `delayed' network collapse is observed in a wide variety of materials \cite{11442,10832,5241,5237,5389,5390,2308,2280,Allain-1364,5926} and seems to be ubiquitous at small $\Uc$. However, while sudden collapse has been attributed to channel formation within the gel \cite{2308,5390}, the microscopic processes operating have never been fully established. A better microscopic understanding of the origin of sudden gel collapse is important not only because the distinctive settling behavior is intriguing from a scientific viewpoint but also because a quantitative prediction of gel stability is a critically important issue in the formulation and manufacture of many commercial products.

The aim of this paper is to report a detailed experimental study of the stability of  gels under gravitational stress. We use a colloidal suspension of nearly monodisperse emulsion drops of radius $\a$ suspended in an index-matched mixture of solvents, which has been well characterized elsewhere \cite{11551}. Gelation is induced by long-range attractive depletion forces. Using time-lapse video imaging we measure the dependence of the height $h$ \nomenclature[h]{$h(\tw)$}{Height of gel at time $\tw$} of a gel upon its age $\tw$, counted from the moment when the gel was formed. No macroscopic sedimentation is observed initially but after a period of latency the gel undergoes a rapid collapse as the system separates into colloid-rich and colloid-poor phases. We investigate the collapse dynamics  as a function of the strength $-\Uc$ of the attractive interactions and the initial height $\h0$ of the gel.  Remarkably we find that when collapse starts the change in the height of the gel $\Delta h = \h0-h$ follows a simple universal dependence on $\tw$ which is independent of the initial height $\h0$ of the gel. The observation of height-independent collapse is surprising and contrasts with the marked height dependence seen in short-range gels studied to date \cite{2308,3708,5386}.  Using a combination of rheology, confocal microscopy and time-lapse video imaging we speculate that the collapse of the gel network occurs as a result of irreversible aging, via a spatially heterogeneous process of localized  `micro-collapses', which leads to a build up of internal stress within the gel and its ultimate failure.

The paper is organized as follows: Section~\ref{sec:materials-and-methods} discusses the preparation of the emulsion gels studied and the experimental techniques used. Section~\ref{sec:experimental-results} details experimental results from both macroscopic and microscopic measurements on the settling behavior of suspensions of attractive particles. The interpretation of the results in terms of internal stress relaxations is discussed in Sec.~\ref{sec:discussion} before we summarize our main findings in Sec.~\ref{sec:summary}.

\section{Materials and Methods}
\label{sec:materials-and-methods}

A long-range attractive interaction was induced between emulsion colloids   by polymer depletion.  The emulsion consisted of poly(dimethyl siloxane) drops dispersed in a solvent mixture of 1,2-ethane diol (ED) and water (mass fraction of ED = 0.59). The solvent composition was adjusted to closely match the refractive index of the emulsion to minimize van der Waals attractions between drops and to enable confocal imaging to be conducted deep within the sample.  A particle radius of $\a = 316 \; \pm \; 11$ nm and a size polydispersity of 0.17 $\pm$ 0.07 was determined from dynamic light scattering measurements.   The thickness of the polymeric stabilizing layer surrounding each emulsion drop was evaluated  by centrifuging a suspension and equating the packing fraction of the sediment to the jamming density of a hard sphere system with the same polydispersity \cite{11210}. This procedure gave a layer thickness of $\approx 7 \pm 1$ nm. The density mismatch between emulsion drops and the continuous phase is $\Delta \rho =  -130 \pm 10$ kg m$^{-3}$. To induce a depletion interaction, we added the non-adsorbing anionic polymer xanthan (Kelco,  $M_{\mys{w}}$ = 4.66 x 10$^{6}$ g mol$^{-1}$). The polymer radius of gyration was determined as $ \rg = 194 \pm 10$ nm by light scattering and viscometry.  The strength of the depletion attraction generated is a function of the polymer concentration and its range is controlled by the relative size $\rg / \a = 0.62 \pm 0.04$ of the polymer and particle. The polymer concentration is quoted here in terms of the dimensionless ratio $\cpstar$, where $c_{\rm{p}}^{*} = 3 M_{\mys{w}} / 4 \pi \rg^{3} N_{\mys{A}} $  is the overlap concentration ($c_{\rm{p}}^{*} =0.25 \conc$) and $N_{\mys{A}}$ is Avogadro's constant. Full details of sample preparation are contained in \citet{11551}.

To monitor the collapse of the gels we used time-lapse video recording to record images of the emulsions as they cream.   A low magnification image of the settling gel was projected onto a CCD camera (Allied Vision Technologies F-080B). A regular sequence of images was captured every 20 seconds. The image series was corrected for optical distortion and non-uniformities in illumination before being calibrated using an accurate grid of lines. The images near the center of the cells were analyzed and the interface separating the upper (dark) phase from the lower (bright) phase was identified automatically using an image analysis routine. The height $h$ of the interface was extracted as a function of time with an accuracy of about $\pm 0.3$ mm. To aid visualization a low concentration ($\approx 0.001 \conc$) of an adsorbing black dye, Sudan black, was added. The dye preferentially partitions into the index-matched PDMS drops so that the colloid-rich phase appears dark in transmitted light.  The colloid-polymer mixtures were thoroughly mixed at the start of the experiments  before being loaded into cylindrical glass vials with an internal diameter of $\d0 = 17$ mm.  The cell diameter was varied between 15 -- 23 mm and both cylindrical glass and poly(styrene) cells were used, with no significant change in collapse behavior. To eliminate air bubbles which lead to irreproducible settling dynamics we used a gentle slow tumbling of the sample vial to thoroughly mix the samples before observation. Repeat experiments showed that following this protocol the collapse kinetics could be measured with a reproducibility of about 10-15\%.

Rheological measurements were performed at 23$\degc$ with a Bohlin HR Nano rheometer (Malvern Instruments). To study simultaneously the temporal evolution of the elastic properties and the height of the gel a novel rheometric vane experiment was developed which allowed  visual observation of the gel as rheological measurements were performed. The vane was made from stainless steel and consisted of four blades (diameter 22.7 mm, height 10 mm). The vane was carefully inserted into clear polycarbonate sample vials (diameter 25 mm, height 65 mm), 10 mm below the top surface of the gel and a thin layer of silicone oil added to minimize evaporation. The vane remained inside the dense upper phase for the duration of the rheological experiments, allowing the collapse process to be continuously monitored. Oscillation measurements were performed at 0.5 Hz at intervals of 200-250~s under controlled stress conditions, within the experimentally-determined linear viscoelastic region, while the height of the gel was monitored simultaneously by time-lapse video microscopy. The absence of wall slip was confirmed by watching the movement of small air bubbles deliberately introduced into samples.

To directly probe changes in the microscopic topology of the gel during collapse we used fluorescent confocal microscopy. The continuous phase of the gel was labeled with $0.02 \conc$ of the fluorescent dye rhodamine-B which combined with the high transparency of the emulsions provided by refractive-index matching allowed high resolution optical visualization deep within the gel. A light microscope (Zeiss, Axioskop S100) was mounted horizontally on its side, at right angles to gravity, and two-dimensional fluorescent images of regions 146 $\times$ 146 $\mu$m  were acquired at 543 nm. The gel was contained in a square cross-section glass vial, with an internal dimension of 13 mm, mounted on a low profile translation stage so that the gel could be imaged at different vertical positions, throughout the full 30 mm height of the sample \cite{11551}. Since the emulsion drop radius is below the optical resolution limit of the confocal microscope we can not identify individual drops. Instead, we concentrated on the larger scale structure of the gel.  The bicontinuous network was identified by thresholding the confocal images to determine the location of the interface separating the (dark) emulsion phase from the (bright) continuous phase. To correct for in-plane variation in the fluorescence yield, each image was divided into 16 sub-images and a local threshold for each sub-image was determined using a cluster-based algorithm \cite{11676}. A careful analysis of the resulting binary images, backed up by direct observation, showed that this approach reliably located the shape and positions of the emulsion and aqueous domains.

\nomenclature[Na]{$N_{\mys{A}}$}{Avogadro's constant}
\nomenclature[Mw]{$M_{\mys{w}}$}{Weight average molecular weight}

\section{Experimental Results}
\label{sec:experimental-results}

\subsection{Collapse dynamics}
\begin{figure}[tbp]%
            \includegraphics[width=0.45\textwidth,viewport=30 110 320 350]{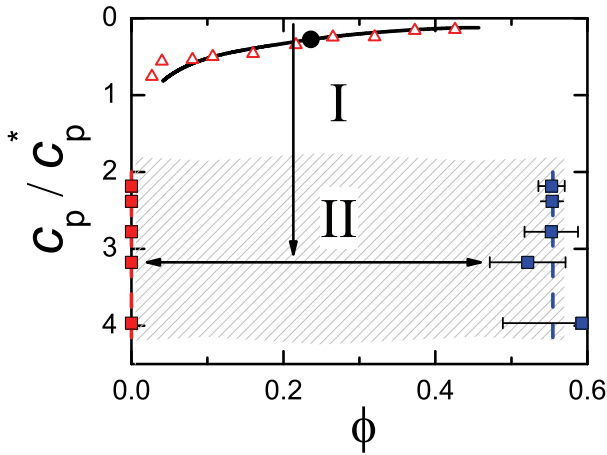}
        \caption{%
        (Color online). State diagram showing gels studied. The solid curve corresponds to the gas-liquid binodal calculated from the generalized free volume theory \cite{8894}, for a polymer-colloid size ratio $\qR = 0.62$. The open triangles identify the experimentally determined phase boundary. The theoretical prediction for the critical point is shown by the filled circle.  The region below the coexistence line can be separated into two kinetic regimes: a region of complete demixing (I), and gelation (II). Quenched into region II, suspensions form a space-spanning network consisting of thick strands of particles. The vertical line indicates the constant colloid volume fraction gels studied ($\phic = 0.213$). The colored symbols represent estimates of $\phicolloid$ for the strands of particles (blue squares) and the coexisting gas (red squares) after phase separation is complete locally. Error bars represent the age-dependent variation in $\phicolloid$ from the same sample. }
        \label{fig:phases}
\end{figure}

To begin, the phase behavior of mixtures of emulsion and polymer was investigated as a function of both the emulsion volume fraction $\phic$ and polymer concentration $\cpstar$. The state diagram plotted in Fig.~\ref{fig:phases} summarizes the results and shows the locations of a stable liquid phase, a narrow region of equilibrium gas-liquid demixing (I) and a broad zone of non-equilibrium gelation (II). The generalized free volume predictions (GFVT) \cite{8894} for a polymer-colloid size ratio $\rg / \a = 0.62$ in a good solvent are shown by the solid lines in Fig.~\ref{fig:phases}. The agreement between the calculated gas-liquid binodal and experiments is good confirming that the experimental system is accurately represented by a simple mixture of hard spheres and non-adsorbing polymer chains.

\begin{figure}[tbp]
    \begin{center}
            \includegraphics[width=0.45\textwidth,viewport=200.0 30.0 720 610]{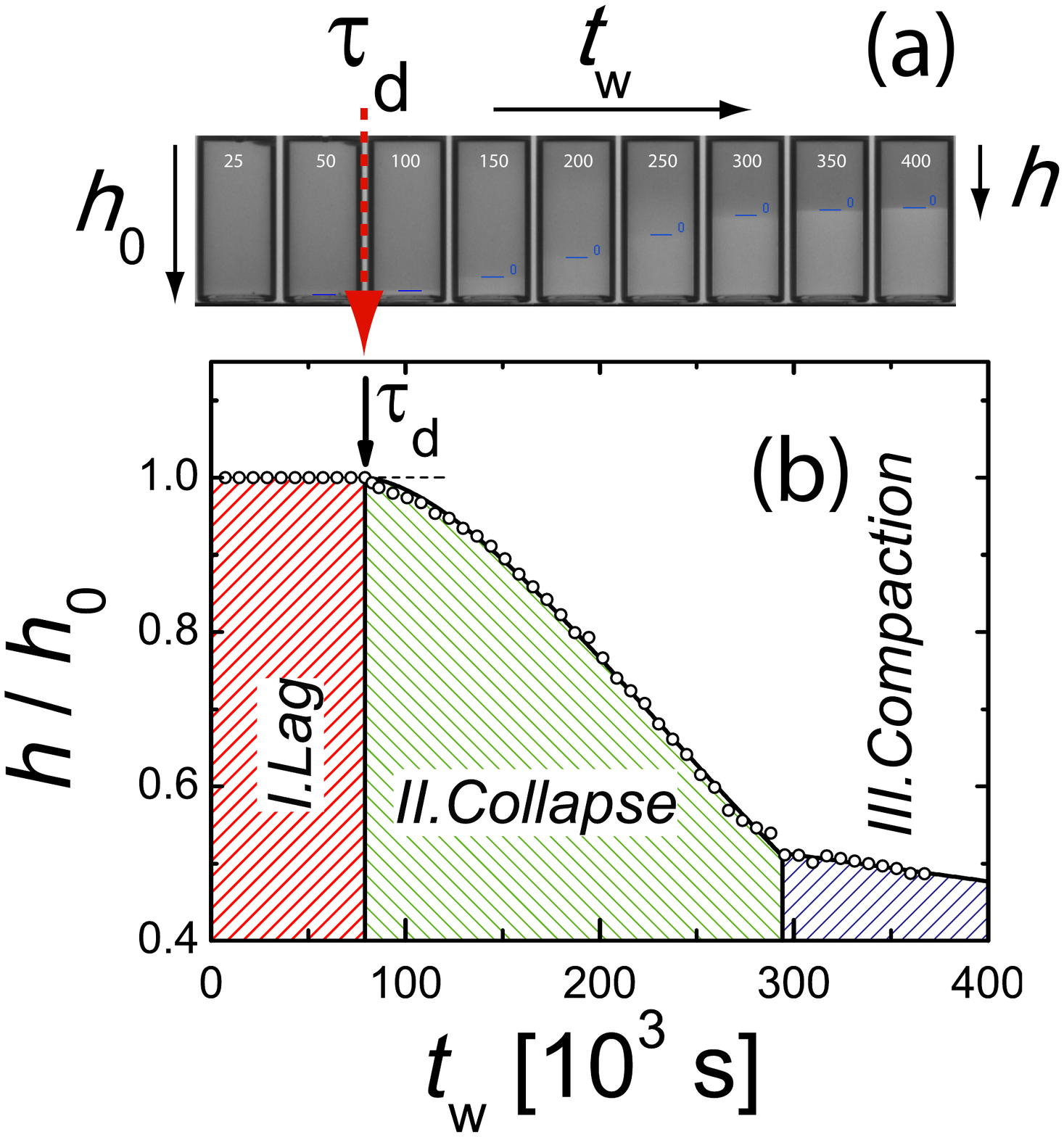}
    \end{center}
    \caption{
    (Color online). Sudden collapse of a gel. (a) Time-lapse images of an emulsion-polymer mixture, with composition $\phic = 0.21$ and $\cpstar = 3.6$, as a function of the time after shaking. Each image is labelled by the time elapsed $\tw$, in units of $10^{3}$ s. The initial height $\h0$ of the sample is 40 mm. The characteristic delay time $\taud$ after which the network starts to collapse is indicated by the dashed arrow.  The solid line denotes the position of the interface between the upper colloid-rich and lower polymer-rich phases. (b) The normalized height $h/\h0$ of the gel shown in part (a) as a function of the elapsed time showing the three stages of settling characteristic of sudden collapse.}
    \label{fig:time-lapse}
\end{figure}

The sudden collapse of gels was investigated as a function of both the strength of the attractions $-\Uc$ and the initial height $\h0$ of the sample. In the absence of polymer, emulsions remained stable and showed no noticeable separation so the mechanical instability seen is a consequence of aggregate formation. The process of collapse is exemplified by the time-lapse CCD images reproduced in Fig.~\ref{fig:time-lapse}(a).  Qualitatively we identify three distinct stages, characterized by the interface velocity $\vc = \mathrm{d}h / \mathrm{d}\tw$, where $\tw$ is the age of the gel. During an initial lag period (I) the network of attractive particles produces a mechanically stable solid, which is capable of supporting its own weight. This regime of solid-like stability persists however only for a limited duration. On times longer than $\taud$, the lag time, the network yields and a clear interface appears (identified by the solid line in Fig.~\ref{fig:time-lapse}(a)). The interface velocity $\vc$ grows smoothly as the gel shrinks and the collapse becomes progressively more rapid. This period (II) of rapid collapse terminates when phase separation nears completion and the interface approaches the final equilibrium plateau. In the final consolidation stage (III), the settling velocity drops markedly as the collapsed gel continues to slowly compress like a solid under its own weight.

\begin{figure}[tbp]
    \begin{center}
    \includegraphics[width=0.45\textwidth,viewport= 100.0 0.0 500 800]{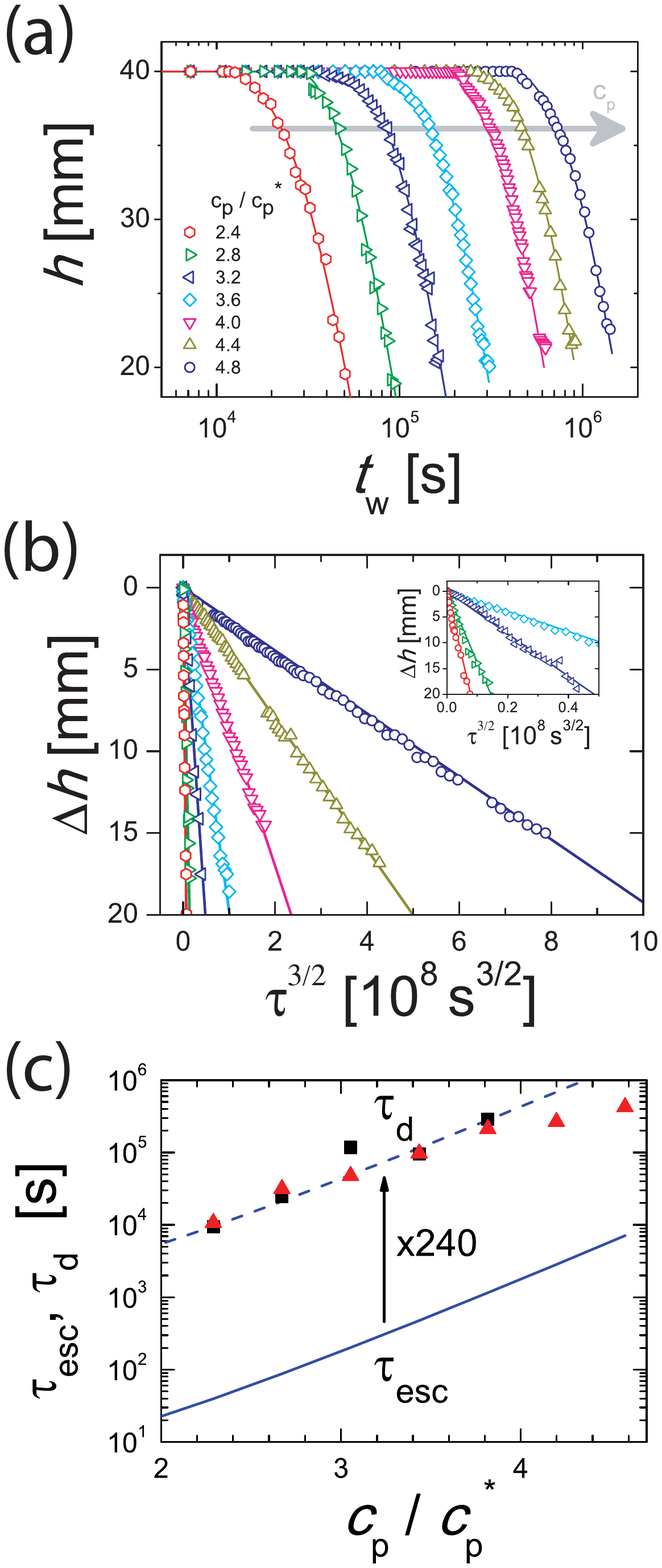}
    \caption{
    (Color online). Gel collapse at fixed height. (a) Time evolution of gel height for different polymer concentrations. Only data in the lag and collapse regimes is shown for clarity.  (b) The change $\Delta h$ in height as a function of $\ts^{3/2}$, where $\ts = \tw - \taud$ is the time elapsed from the start of collapse. Symbols and data are the same as in (a). The solid lines are straight line fits. The inset shows an expanded view of the short-time data. (c) Comparison between the experimentally-measured lag times $\taud$ (filled points) and the average lifetime $\tauesc$ of an individual particle bond (solid line). Lag times were measured  in both glass (triangles) and poly(styrene) vials (squares). The dashed line, which reproduces the experimental data reasonably well, equates to a fixed number of bond lifetimes ($\taud \approx 240 \; \tauesc$).}
    \label{fig:collapse-dynamics}
    \end{center}
\end{figure}

The lag time $\taud$ is a strong function of the polymer concentration and hence the strength of depletion attractions. Fig~\ref{fig:collapse-dynamics}(a) shows the time-dependent height $h(\tw)$ of gels prepared with different polymer concentrations but for the fixed initial height $\h0 = 40$ mm. Inspection
reveals two striking features. First, as reported in previous work \cite{5389,5237}, $\taud$ grows strongly with increasing polymer concentration. Indeed the concentration dependence of the lag time is well described by the exponential relationship, $\taud \sim \exp ( \cpstar)$, as shown in Fig.~\ref{fig:collapse-dynamics}(c). Second, the sedimentation profiles are remarkably similar in shape when plotted in a linear-log representation. The height profile at a low polymer concentration may be simply mapped onto a high concentration sample by shifting the collapse profile to the right along the logarithmic time-axis. To explore this scaling behavior quantitatively, we focus purely on the collapse regime and replot the change $\Delta h = \h0-h$ in the height of the gel as a function of the shifted time variable $\ts = \tw - \taud$, the time counted from the instant when the network first yields. Fig.~\ref{fig:collapse-dynamics}(b) demonstrates, rather unexpectedly,  that the initial change in the height of the gel depends linearly upon $\ts^{3/2}$, over a wide range of polymer concentrations. In the early stages of gel collapse, where $\Delta h \lesssim 0.5 \h0$, the height of the gel follows approximately the algebraic expression
 \begin{equation}\label{eqn-experimental-t3/2}
    h(\tau) = \h0 - A \tau^{\frac{3}{2}}
 \end{equation}
 where $A$ is a polymer-dependent prefactor. The existence of this simple relationship suggests a single common mechanism is controlling the collapse of the gel at different polymer concentrations. We return to a more detailed discussion of this point in Sec.~\ref{sec:discussion} to speculate on a possible origin of the $\tau^{\frac{3}{2}}$ scaling found here. Finally, we note that an alternative scaling has been suggested  by Kilfoil et al. \cite{2905}. We found however that their approach failed when applied to the wide range of polymer concentrations and heights studied here.

\begin{figure}[tbp]
    \begin{center}
        \includegraphics[width=0.45\textwidth,viewport=100 200 500 750]{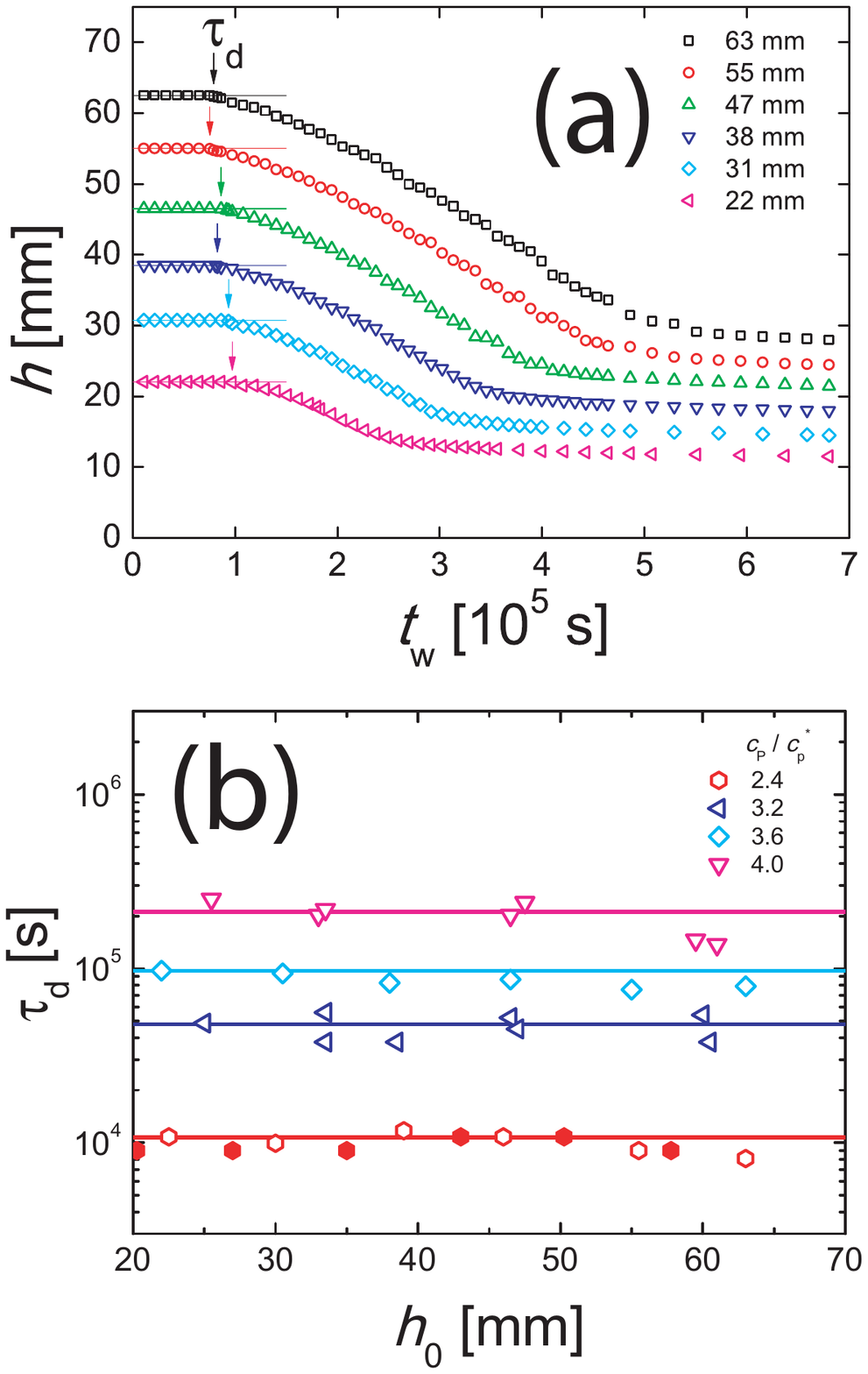}
    \end{center}
    \caption{%
        (Color online). Height-independent lag times. (a) Temporal evolution of a gel with different  initial heights but a fixed composition, $\phic = 0.213$, and $\cpstar = 3.6$. The arrows indicate the lag time $\taud$.  (b) The lag time $\taud$ as a function of the initial height $\h0$ of the gel, for a number of different polymer concentrations. The open symbols denote data obtained in glass-walled cells while the filled symbols indicate measurements in poly(styrene) cells. The nature of the cell wall has no noticeable effect on the delay time measured.
     }%
     \label{fig:Height-independent-delay-times}
\end{figure}

Previous studies \cite{2308,5243} of gel collapse have suggested that the mechanism  of collapse depends sensitively on the initial height $\h0$ of the gel. Gels formed in short sample cells display steady or `creeping' sedimentation where the height falls continuously with age at a rate which decays exponentially with time while taller samples show sudden collapse. To test whether this behavior is intrinsic to the long-range systems studied here we have varied the initial height $\h0$ and monitored the evolution of the height of the gel with time. In all the samples reported here, polymer concentration from $\cpstar = 2.4 - 4.0$ and heights $\h0 = 22 - 63$ mm,  sudden gel collapse was always observed and we saw no transition to creeping sedimentation. Fig.~\ref{fig:Height-independent-delay-times}(a) shows a representative set of data where the time evolution of the interface height $h(\tw)$ is plotted for a range of initial heights and a single polymer concentration ($\cpstar = 3.6$). Inspection of the data reveals that rather surprisingly the time $\taud$ during which the gel is solid-like is largely independent of the initial height $\h0$ of the sample. Measurement of the variation of $\taud$ with height for a wide range of polymer concentrations, presented in Fig.~\ref{fig:Height-independent-delay-times}(b), confirms this observation. We have checked that this is not due to solid friction between the gel and cell wall \cite{5386} by repeating measurements at $\cpstar = 2.4$ using cylindrical poly(styrene) cells to alter the degree of wall adhesion. The results, shown as the filled data points in Fig.~\ref{fig:Height-independent-delay-times}(b), are in excellent agreement with the data obtained in the glass vials (open points), demonstrating that wall friction is unimportant.

Since the gel is initially a solid, the top of the sample vial is subject to a gravitational stress $\sigg$, which is generated by the full buoyant stress of all of the suspension below, so $\sigg = \Delta \rho g \phic \h0$. \nomenclature[g]{$g$}{Acceleration due to gravity}  Taking $\phic = 0.21$ and $\Delta \rho = -130$ kg m$^{-3}$ we estimate that $\sigg$ varies from between approximately 5 to 15 Pa for the heights used here. The values for the buoyant stress considerably exceeds the yield stress of the gel network, which we estimate from rheological measurements as $\sigy \sim 0.1$ Pa, so even while $\sigg$ is an order of magnitude larger than the stress required to break the gel's load-bearing structure the gel does not collapse macroscopically. The insensitivity of $\taud$ to $\h0$ means we can rule out the possibility that the initial period of latency of the gel is determined purely by the breaking of single uncorrelated bonds. Collapse clearly requires a substantially larger degree of restructuring of the network than
is necessary simply for mechanical yielding.

\begin{figure*}[tbp]

    \begin{center}
            \includegraphics[bb=50 230 595.22 600,width=0.9\textwidth]{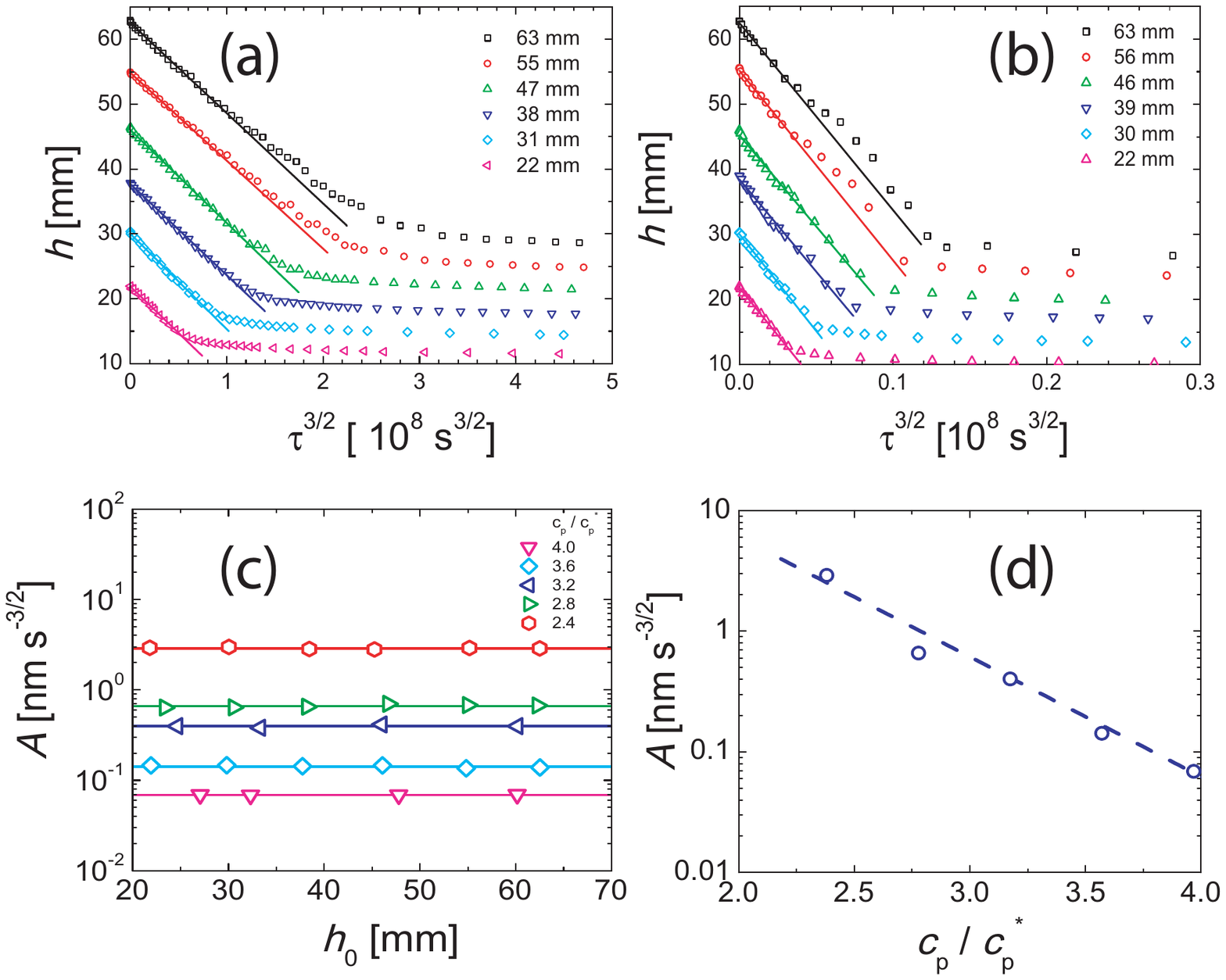}

    \end{center}
    \caption{%
        (Color online). Initial $\ts^{3/2}$--collapse dynamics. (a) The height $h$ of a gel plotted as a function of the $3/2$--power of the time elapsed after the gel yields, for different initial heights $\h0$. Curves are labelled by the initial height. The samples have fixed polymer concentration $\cpstar = 3.6$, and colloid content $\phic = 0.213$.  (b) Similar time-dependent settling observed in gels with  $\cpstar = 2.4$. (c) Invariance of the prefactor $A = -\lim_{\ts \rightarrow 0} \mathrm{d}h /\mathrm{d}\ts^{3/2}$ with the height of the gel. Curves are labelled by $\cpstar$. (d) Exponential dependence of $A$ on the polymer concentration, highlighting the activated nature of the collapse process.
     }%
     \label{fig:t3/2-plots}
\end{figure*}

 To explore the effect of height on the kinetics of collapse we focus on the initial rate of collapse of the gel with different $\h0$.  Figure~\ref{fig:t3/2-plots} shows that the power-law expression (Eq.~\ref{eqn-experimental-t3/2}) which captures well the height variation in gels with a fixed $\h0$ also holds for gels with a wide variety of different starting heights.  The gel does not collapse with a fixed time-invariant velocity but rather the interface velocity $\vc = \mathrm{d}h / \mathrm{d}\tau$  behaves at short times like $\tau^{1/2}$, a behavior which hints at a surprisingly novel mechanism of collapse. To further investigate this mechanism we have studied the dependence of the prefactor $A = \mathrm{d} \Delta h / \mathrm{d} \tau^{3/2}$  on the height and polymer content of the gel. Fig.~\ref{fig:t3/2-plots}(c) shows that rather remarkably, for all the gels studied, the prefactor $A$ does not change with the initial height of the gel. Since the gravitational stress $\sigg$ on the gel increases linearly with its height, this result  suggests that  $\sigg$ is relatively unimportant in determining the process of collapse, at least under the conditions of our experiments. The central role of thermally-induced bond dissociation is seen in Fig.~\ref{fig:t3/2-plots}(d) where it is revealed that the prefactor $A$ scales exponentially with $\cp$,  equivalent to an exponential dependence on the depth of the interaction potential.  In conclusion, the process of collapse appears to be thermally rather than stress-activated.


%
%

 To interpret these striking observations we model the initial deformation of a gel using the poroelastic formalism first introduced by Buscall and White \cite{5922}. In this approach the gel is treated as a biphasic fluid-saturated porous continuum with the pore pressure $P$ as a state variable. A gel consists of two distinct phases: a solid phase of connected strands of emulsion particles, and a second liquid phase consisting of a fluid solution of a non-adsorbing polymer. In the early stages of collapse the gel is essentially uncompressed so there is no elastic stress due to deformation and the rate of collapse is limited essentially by the rate at which fluid is forced out of the gel \cite{3708}. Defining $v$ as the macroscopic velocity of the fluid flow through the gel and $w$ as the local displacement of the solid network along the gravitational $z$-axis then using Darcy's law,
 \begin{equation}\label{eqn-darcy}
      - \partial_{z} P = \frac{ \eta(1-\phicolloid)}{k} (v-\partial_{t} w),
\end{equation}
where $k$ is the permeability of the network, which since we are considering only the early stage of collapse we assume to be height independent, and $\eta$ is the viscosity of the continuous phase. Continuity demands
\begin{equation}\label{eqn-continuity}
    (1-\phicolloid) v = -\phicolloid \partial_{t} w
\end{equation}
which, since $\phicolloid < 1- \phicolloid$, implies that the fluid velocity $|v|$ must be small \cite{3708} in comparison to $|\partial_{t} w|$. Consequently, if the displacement of the gel varies as $\ts^{3/2}$ then, from Eq.~\ref{eqn-darcy}, the pressure gradient at the top of the gel, adjacent to the interface, must be increasing  as $\ts^{1/2}$. This time dependence rules out a simple compression of the gel as a consequence of gravity because the pressure gradient would then be a constant, $\partial_{z} P = - \Delta \rho g \phic$, and the gel would accordingly shrink linearly with time \cite{3708}. The $\ts^{1/2}$ dependence of $P$ suggests instead a diffusive process may be responsible for the unusual collapse dynamics seen. We can however rule out a bulk diffusive process of conventional syneresis, akin to the shrinkage of a polymer gel undergoing a phase transition \cite{9558}. In this case, the contraction of the matrix as the phase separation ensues would generate an increase in $P$ which leads to the expulsion of fluid and a shrinkage of the gel as $t^{1/2}$ \cite{11741} rather than the $t^{3/2}$ dependence seen here. Furthermore, this process would require fluid to be transported through the full length of the system so the rate of diffusion would depend on the height of the gel, which is also incompatible with our data.  Clearly, a new mechanism is required to correctly explain the observed data.

\subsection{Microstructure}
\label{sec:microstructure}

\begin{figure}[tbp]
            \includegraphics[width=0.45\textwidth,viewport = 25 35 730 530]{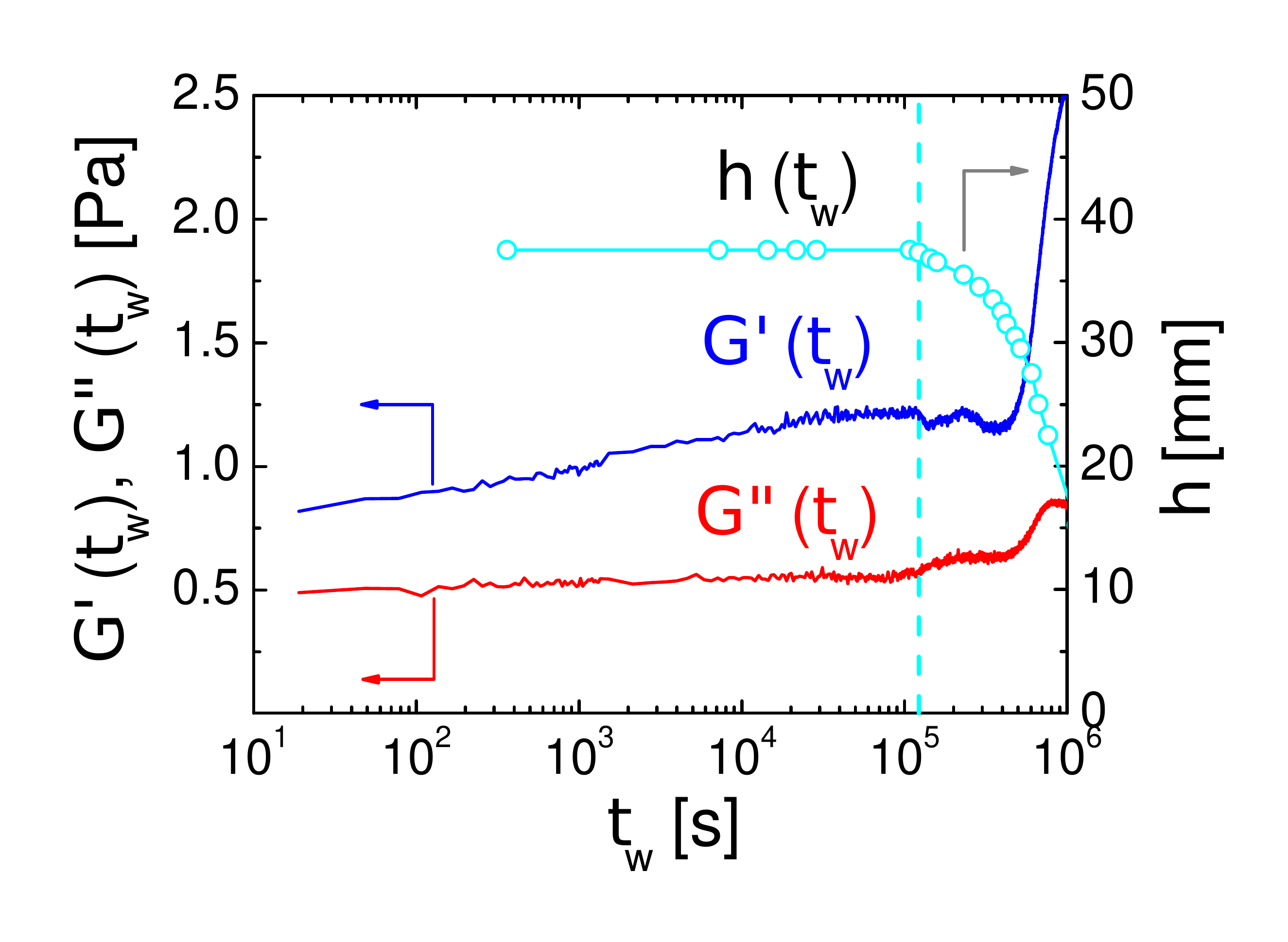}

%
    \caption{(Color online). Simultaneous measurements of linear viscoelasticity and height $h$ of gel as a function of time elapsed  since preparation $\tw$.  Gel had an initial composition of $\cpstar = 3.6$, $\phic = 0.213$ and a measured delay time of $\taud = 1.2 \times 10^{5}$ s (indicated by dashed line). The elastic $G'$ and loss $G''$ moduli were measured by applying a oscillatory stress of magnitude 0.0025 Pa at a frequency of 0.5 Hz and measuring the strain response. The gel stiffens continuously with age until the elastic modulus drops at $\tw = \taud$ as the gel begins to collapse.}
    \label{fig:vane-rheology}
\end{figure}

Sudden collapse reflects a dramatic loss of mechanical integrity as the gel ages. To understand the nature of this mechanical failure, we first examine the evolution of the elastic properties of the gel while simultaneously recording the height of the gel. Figure~\ref{fig:vane-rheology} shows the rheology and interface height $h(\tw)$ during the settling of a representative gel. At the earliest times recorded, the sample is solid-like with an elastic shear modulus $G'$ which is larger than the viscous modulus $G''$. The measurements however reveal that the elasticity of the gel far from reducing with time, as one might naively expect, actually increases continuously up to the point $\tw = \taud$ when the gel starts to collapse. Immediately collapse starts, $G'$ also drops, but only by a relatively small amount (less than 10\%), before finally growing steadily as the emulsion volume fraction in the upper phase increases with the onset of phase separation. We see no sign of large-scale hydrodynamic mixing, recirculation, and the development of channels which have been seen in some other studies of gravitational collapse \cite{2308,5390,10832}.
It is intriguing that even though $G'$ drops at the initiation of collapse the overall mechanical response remains predominantly elastic ($G' > G''$). We suspect that this reflects the inhomogeneous nature of the network at collapse. If the sample is heterogeneous the  response measured will be an average over regions where the gel has broken apart, and is a fluid, with other portions of the network which remain elastic. The recorded response will then depend on the relative sizes of the mechanical vane and the inhomogeneous regions within the gel. To confirm that the sample becomes a fluid when the gel begins to collapse we have placed a small glass block about half the width of the cell in the base of the cell. When the gel begins to collapse the interface between the gel and the bottom of the cell rapidly flattens indicating that the base of the gel becomes a fluid of aggregates as the gel collapses.
%
%
%
%
%
%

\begin{figure}[tbp]
    \begin{center}
       \includegraphics[width=0.45\textwidth]{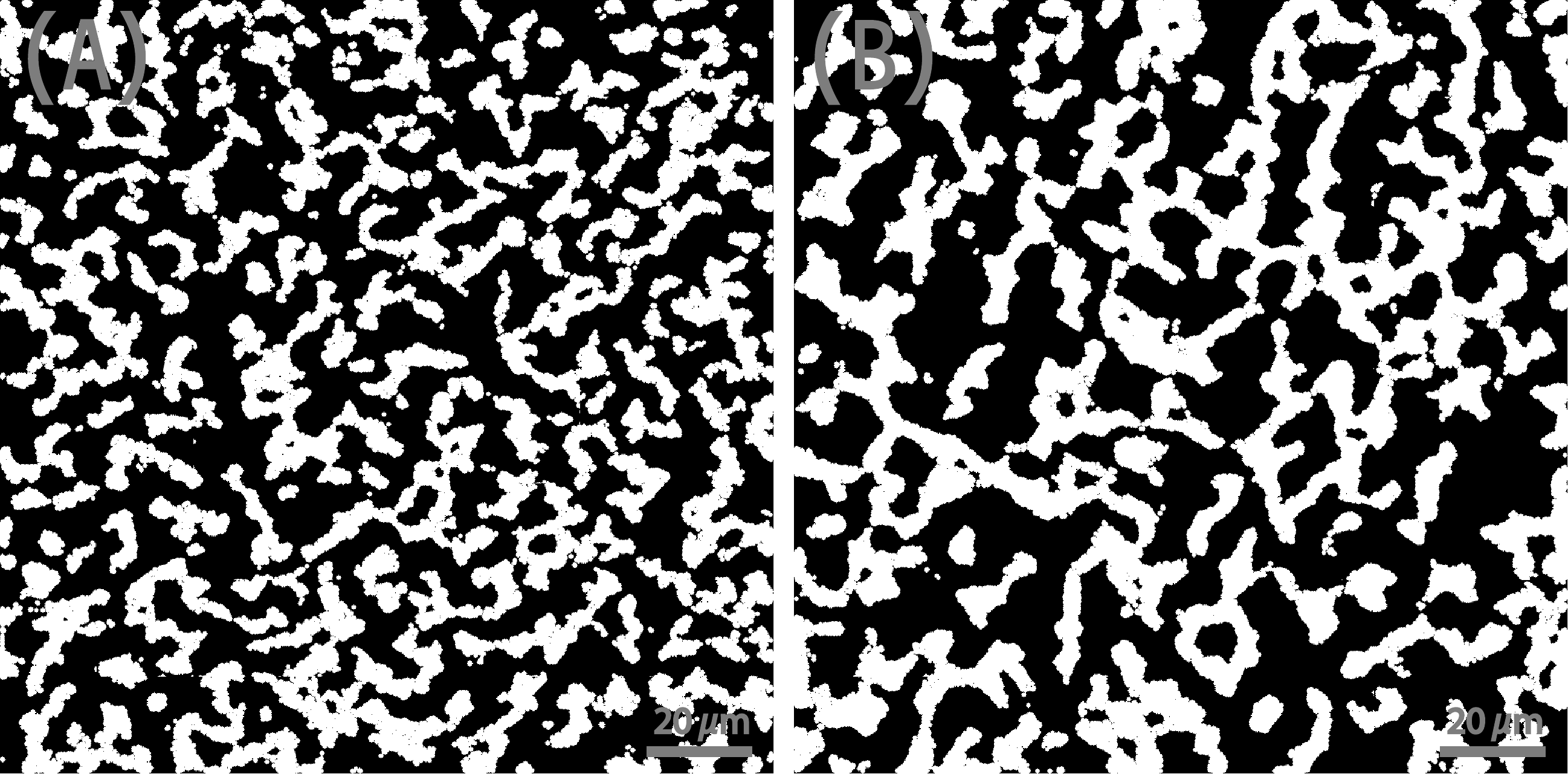}
     \caption{Continuous coarsening of gel with age. Two-dimensional binary representation of a gel with composition $\cpstar = 2.4$, $\phic = 0.213$ after (a) $3.6 \times 10^{3}$ s, and (b) $1.12 \times 10^{5}$~s from preparation. Gravity points vertically downwards and the scale bar corresponds to 20~$\mu$m.}
      \label{fig:aging}
\end{center}
\end{figure}

While the rheological measurements provide a mechanical insight into gel settling, they do not clarify the link between the macroscopic processes of collapse and the microscopic structural reorganization occurring during aging and sedimentation. Indeed, at first sight, it seems counter-intuitive that a gel which is becoming gradually stiffer with time should ever collapse at all. To probe the link between the microscopic and macroscopic length scales, we have examined the temporal evolution of the gel microstructure using confocal microscopy. The coarsening is illustrated by the binary two-dimensional images reproduced in Fig.~\ref{fig:aging}. In the rectangular cell used for imaging experiments, the delay time was measured as $\taud \approx 1.3 \times 10^{5}$ s so both images refer to the latency period before collapse starts. Clearly, although the gel remains mechanically stable during this period there is a slow but continuous evolution in the microscopic  nature of the particle network and the system is not arrested. A closer look at the data in Fig~\ref{fig:aging} reveals that the interface between the continuous and particle phases is quite rough, suggesting that surface tension is unimportant and the dense portion of the gel is not a fluid. Direct observation show that particle diffusion is strongly suppressed and particles move infrequently between the strands of the network, indicating that the interaction network is probably glassy. The images in Fig~\ref{fig:aging} illustrate two further microstructural characteristics which will be important to our later discussion on the mechanism of network collapse. First, it is evident that the thickness of the backbone of the gel grows slowly but continuously with time, a fact which probably explains the increase of $G'$ with $\tw$ seen in Fig.~\ref{fig:vane-rheology}. Second, the thickness of the network of particle strands is not uniform. The gel contains a relatively large number of thin junction points where two or three arms (in 2D) are connected together.  Simultaneous breakage of the particle bonds at these relatively weak junction points would lead to a rapid break up of the whole network.

\nomenclature[N1]{$N(l)$}{Number of chords with length between $l$ and $l+\mathrm{d}l$ in the dense phase}
\nomenclature[N0]{$N$}{The total number of chords in the dense phase}

To quantify the change of the microstructure with time, we use chord methods developed to analyze statistically random heterogeneous materials \cite{11647}. We superimpose on the images of Fig~\ref{fig:aging} a uniform grid of horizontal and vertical lines. The two-phase interfaces divide each grid line into chords that are either inside the dense part of the network or else lie within the solvent background. We define a chord as the line segment between two consecutive intersections of the random line with the bounding two-phase interface. Focusing only on those chords that lie \textit{within} the dense strands of the network, we count the number of chords $N(l)$ with lengths in the range $l$ and $l+\mathrm{d}l$. If $N$ is the total number of chords then the degree of linear `connectedness' of the gel may be characterized in terms of a probability density function, $p(l)$, where \nomenclature[p]{$p(l,\tw)$}{Chord-length distribution function}
\begin{equation}\label{eqn-chord-prob-den-func}
    p(l)\mathrm{d}l = \frac{N(l)}{N}.
\end{equation}
The quantity $p(l)\mathrm{d}l$ is the probability that a randomly-chosen chord has a length between $l$ \nomenclature[l]{$l$}{Chord-length} and $l+\mathrm{d}l$. We find that the chord length distribution $p(l)$ displays a characteristic shape, with $p(l)$ first increasing with growing $l$ before reaching a maximum at a finite $l$ and finally decaying exponentially for larger $l$. The diameter of the strands of particles within the gel may be characterized either from the value at which $p(l)$ takes a maximum value or from the mean chord length $\lg$
\begin{equation}\label{eqn-mean-chord-length}
    \lg = \int_{0}^{\infty} l p(l) \mathrm{d}l ,
\end{equation}
which we use here because it displays a smaller statistical error. \nomenclature[lg]{$\lg$}{Mean chord length in gel phase \verb+\lg+} Finally, the chord functions also provide an efficient means to estimate the volume fraction of the high density colloidal regions in the gel. If we assume the gel is isotropic and the two-dimensional images are chosen randomly then the fraction $\chi$ of the volume of the gel occupied by the dense regions  is  \nomenclature[\chi]{$\chi$}{Fraction of volume of gel occupied by dense phase}
\begin{equation}\label{eqn-vol}
    \chi = \frac{1}{L}\sum_{l} N(l) l
\end{equation}
where $L$ is the total length of the original lines. \nomenclature[L]{$L$}{length of original line} The volume fraction of colloids, $\phig$, in the dense regions of the gel is given by the ratio $\phic / \chi$, where $\phic$ is the initial colloid volume fraction.

\begin{figure}[tbp]
    \begin{center}
        \includegraphics[width=0.45\textwidth,viewport=50 0 500 700]{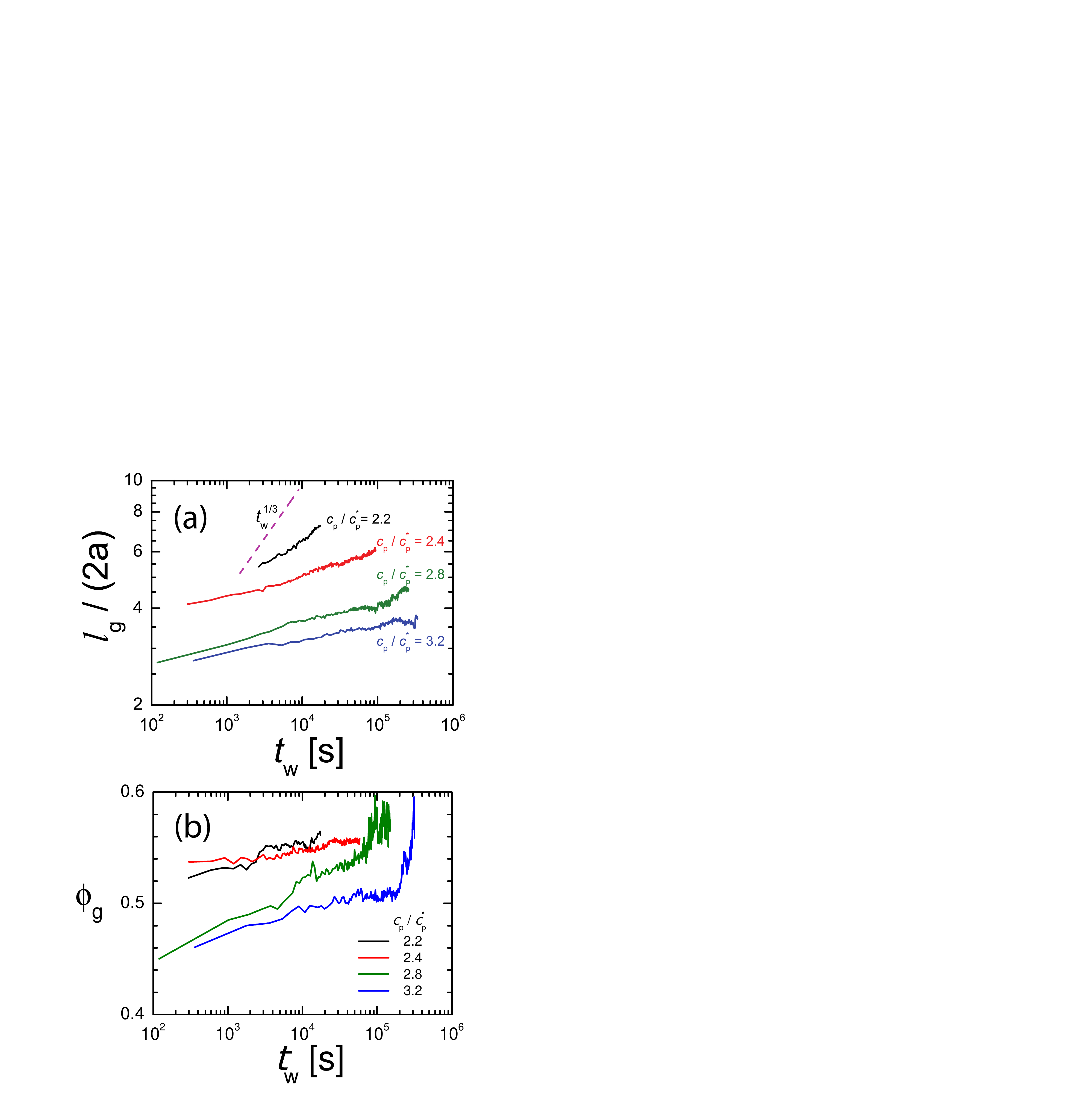}
    \caption{%
        (Color online). Evolution of colloidal network with time. (a) Average chord length $\lg$ in the colloid phase, in units of the particle diameter. The rate of increase of $\lg$ is always slower than the diffusive growth, $\lg \sim \tw^{1/3}$ (shown by dashed line) characteristic of classical spinodal decomposition \cite{11738}, and slows down considerably with increasing polymer concentration $\cpstar$. (b) Average colloid volume fraction $\phig$ within dense regions of the gel, as a function of elapsed time. The density of the particle strands within the network increases progressively with age before rising rapidly at the onset of gel collapse.
     }%
     \label{fig:Evolution-of-colloidal-network}
  \end{center}
\end{figure}

Next we characterize the slow evolution of the gel structure. Fig.~\ref{fig:Evolution-of-colloidal-network}  shows the age dependence of the mean chord length $\lg$ and the volume fraction $\phig$ prior to collapse. A key observation is that during the latency period when the height of the gel is unchanged structural reorganization is never fully arrested but continues, albeit rather slowly. So, for instance, the growth of the mean chord length is well described by a power law, $\lg \sim \tw^{\alp}$, with an exponent $\alp$. \nomenclature[\alpha]{$\alp$}{The growth exponent \verb+\alp+} The growth law is always much slower than the $\tw^{1/3}$ dependence expected for the diffusive regime of classical liquid-gas phase separation. The growth exponent $\alp$ depend rather strongly on the polymer concentration with $\alp$ decreasing markedly as $\cpstar$ is increased. Similar slow growth has been identified previously in simulations of deep quenched Lennard-Jones systems \cite{11738}  and has been interpreted as indicating that the dense domains of the gel are actually glassy.


\subsection{Origin of lag time}
\label{sec:origin-delay}

Having characterized the macroscopic process of collapse, we now discuss the mechanism for the initial failure of the particle network. A gel is a metastable phase with a high free energy density whose consolidation is driven ultimately by the thermodynamic driving force for phase separation. However once a stable percolating network of strands of particles is formed the dynamics of phase separation is slowed down considerably because, as evident from Fig.~\ref{fig:aging}, the strands of the network are many particles wide so large scale rearrangements of the gel require the simultaneous dissociation of all of the particle bonds within the cross-section of a strand, which will be very rare. The network accordingly lowers its free energy via a series of small-scale structural reorganizations which proceed through the rupture of essentially single particle bonds.  The breakup of an energetic bond between particles, diffusion to dense region of the network, and a reformation of the broken bond allows a net increase in the number of nearest neigbouring particles with a concomitant lowering of the free energy of the system. For the network to coarsen, the system must overcome the energetic barrier associated with single bond rupture. This could be achieved, in principle, either thermally or as a result of an applied stress. The observation that the delay time is unaffected by the initial height of the gel strongly suggests  that the delay time and hence the rupture of individual bonds is controlled primarily by thermal fluctuations rather than being stress-driven.

\begin{figure*}[tbp]
    \includegraphics[width=0.9\textwidth,viewport= 40 250 540 630 ]{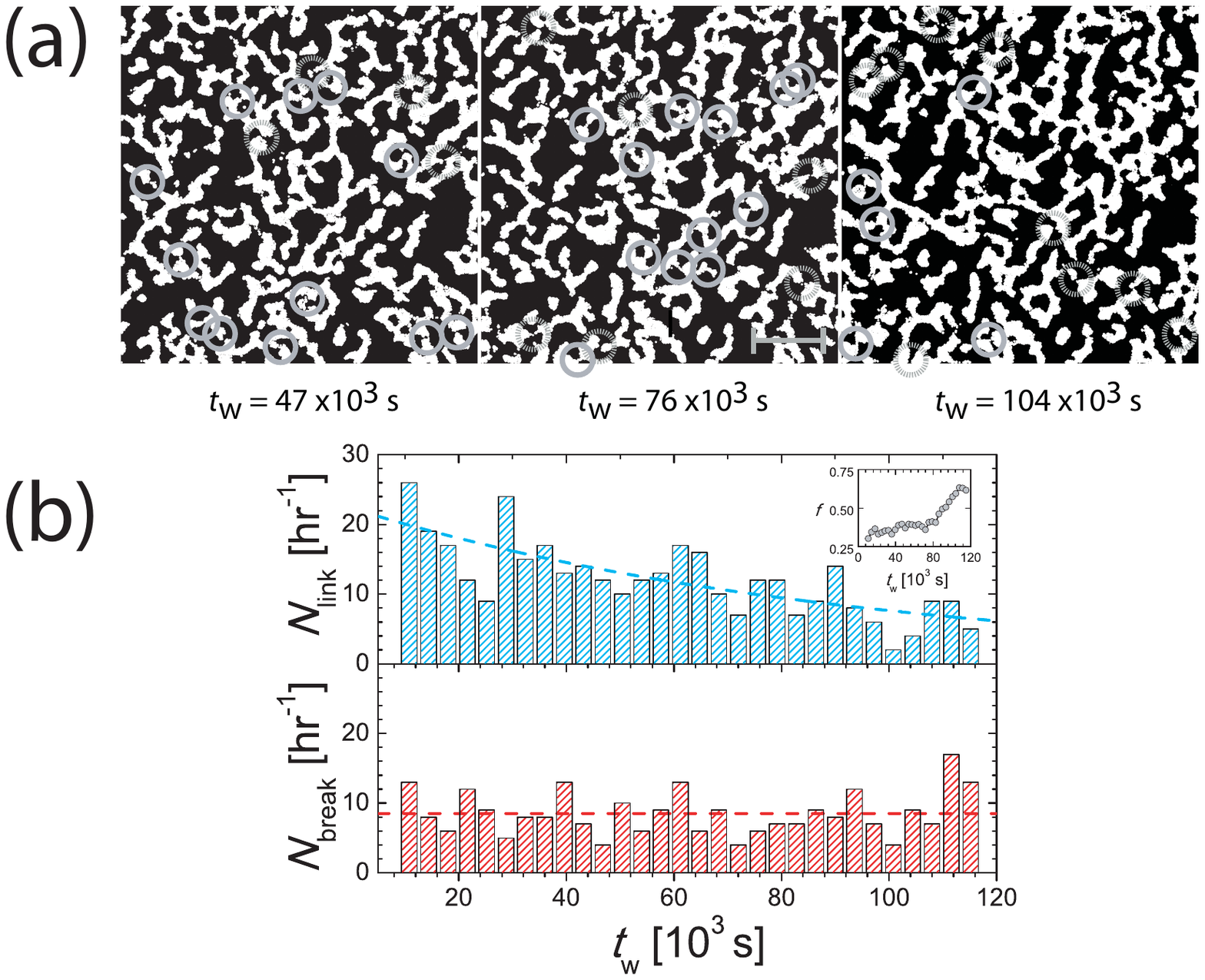}
    \begin{center}

%
    \end{center}
    \caption{%
        (Color online). (a) 2D confocal images of coarsening gel network formed at $\phic = 0.213$, and $\cpstar = 2.4$. The particles are shown in white. The solid circles indicate network junctions which have formed in the preceding 3600 s. The dashed circles indicate positions where the network has, in the same interval, broken. The scale bar is 30 $\mu$m long. Network collapse occurs at $\taud \sim 120 \times 10^{3}$ s. (b)  The number of reassociation $\nlink$ and rupture $\nbreak$ events per hour as a function of the age $\tw$ of the gel. The dashed lines are guides to the eye. The inset shows the $\tw$-dependence of the fraction $f$ of rupture events.
     }%
     \label{fig:aging-events}
\end{figure*}

To calculate the average lifetime $\tauesc$ \nomenclature[\tausec]{$\tauesc$}{Kramers escape time \verb+\tauesc+} of an individual particle bond due to thermal fluctuations, we assume that a single bond ruptures on a scale comparable to the time it takes a Brownian particle to escape from an attractive ramp potential with the same range $\D$ and depth $-\Uc$ as the interparticle depletion potential. The mean escape time in the overdamped limit is given by the Kramers expression \cite{10855}
\begin{equation}\label{eqn-Kramers}
    \tauesc = \frac{\delta^{2}}{\Dt} \frac{\exp (-\Uc) - (1-\Uc)}{(\Uc)^{2}}
\end{equation}
where $\Dt$ is a translational diffusion constant. \nomenclature[Dt]{$\Dt$}{Translational diffusion constant \verb+\Dt+} We estimate $\Dt$ from the short-time self diffusion constant in a hard sphere suspension at the same $\phicolloid$,    which since the dense regions of the gel have a volume fraction $\approx 0.55$  is about 20\% of the dilute free particle limit, $\Diff = \kb T/(6 \pi \etaL \a)$. \nomenclature[D0]{$\Diff$}{Single particle diffusion constant \verb+\Diff+} \nomenclature[\etaL]{$\etaL$}{Low shear limiting Viscosity \verb+\etaL+}  The limiting low shear viscosity $\etaL$ was determined by extrapolating measurements of the steady-shear rheology of the polymer solution to a vanishing shear rate and fitting to the Martin equation,
\begin{equation}\label{eqn-Martin}
    \frac{\etaL}{\vis0} = 1 + [\eta]\cp \exp \left( k_{\mys{H}} [\eta]\cp   \right)
\end{equation}
which has been found to correlate well viscosity in dilute and semi-dilute polymer solutions ($\cpstar < 10$). Here $[\eta]$ is the intrinsic viscosity, $\vis0$ is the viscosity of the mixed solvent, $\cp$ the polymer mass concentration, and $k_{\mys{H}}$ is a constant (equivalent to the Huggins constant at low $\cp$). Fitting data in the range $\cp = 0.6-1.2 \conc$ to Eq.~\ref{eqn-Martin} gave $[\eta] = 2.32$ ml/mg and $k_{\mys{H}} = 1.2$. The width of the depletion zone $\delta$ and the potential at contact $-\Uc$ were estimated using the generalized free volume theory for mixtures of hard sphere colloids and excluded-volume polymer chains, as detailed in Ref.~\cite{11551}. Figure~\ref{fig:collapse-dynamics}(c) shows a  comparison between the measured delay time $\taud$ and the average lifetime $\tauesc$ of a single particle bond, estimated from Eq.~\ref{eqn-Kramers}. The ratio of the two timescales is very nearly constant, for a wide range of polymer concentrations, with the delay time approximately 240 times the estimated single particle Kramers escape time. This strong correlation highlights the pivotal role of spontaneous thermal fluctuations in determining
the latency period of the gel. The fact that the delay time is many times longer than the rupture of a single bond probably reflects the cooperative nature of gel failure. The strands of the network are several particles wide so failure requires the simultaneous dissociation of all of the bonds in the cross-section of a particle chain \cite{11877}. The alternative picture proposed by Buscall et al. \cite{9091}, that the ratio  $\taud / \tauesc$ is determined by the mean coordination number of particles within the gel, could only be consistent with our observations if the mean particle coordination number varied with the depth of the attractive potential. To distinguish completely between these two possibilities requires a more detailed microscopic model of gel failure than is currently available.

\nomenclature[\eta0]{$\vis0$}{Viscosity of solvent \verb+\vis0+}
\nomenclature[\eta]{$\eta$}{Viscosity}
\nomenclature[kH]{$k_{\mys{H}}$}{Martin equation constant}

To explore the microscopic mechanism by which thermal fluctuations lead to delayed failure we used real-space confocal imaging to follow the time evolution of the gel. Since the load-bearing nature of the network is clearly important we concentrate on changes in the connectivity of the strands of particles which constitute  the gel. Figure~\ref{fig:aging-events}(a) shows two-dimensional confocal slices taken from the same physical region within an aging gel before  collapse occurs. Because of the finite bond energy, the network structure slowly but continuously evolves, with fluctuations in both the number and type of junction points. By comparing 2D images of the fine-stranded structure of the network at hourly intervals we identified discrete strand association and dissociation events occurring over this period. Examples where the strand network is ruptured are indicated by the dashed circles in Fig.~\ref{fig:aging-events}(a) while the solid circles identify new cross-links formed by the reassociation of strands. Counting the number of reassociation $\nlink$ and rupture $\nbreak$ events recorded per hour, as a function of the age of the gel, results in the data shown in Fig.~\ref{fig:aging-events}(b). There is gradual reduction over time in the number of reassociation events $\nlink$, as the network is formed in an open high-energy state and then relaxes slowly into a lower more compact structure. Strikingly however, we see that the rate of bond rupture does not show the same slowing-down. $\nbreak$  is essentially independent of age, presumably because rupture is an activated process which is dominated by the single particle bond energy barrier. The consequence of the different time dependence seen for association and rupture is that the proportion of breakage events $f = \nbreak/(\nbreak+\nlink)$ (shown in the inset of Fig.~\ref{fig:aging-events}(b)) grows with the age of the gel. The increasing proportion of strand ruptures ultimately leads to failure of the stress-bearing backbone of the gel and the initiation of collapse.

\subsection{Appearance of structural heterogeneities}
\label{sec:heterogeneous}

Work in the last decade \cite{8739,4480} has shown that soft glassy materials frequently display structural heterogeneities. In materials where the elastic behavior of a material dominates over its viscous response any deformation due to a local rearrangement can propagate macroscopic distances so the size of regions which undergo correlated rearrangements can be sizeable. If this holds true in our system, then it should be feasible to see signs of the long-range distortion field generated by local rearrangement events by microscopy.

\begin{figure}
        \includegraphics[width=0.45\textwidth]{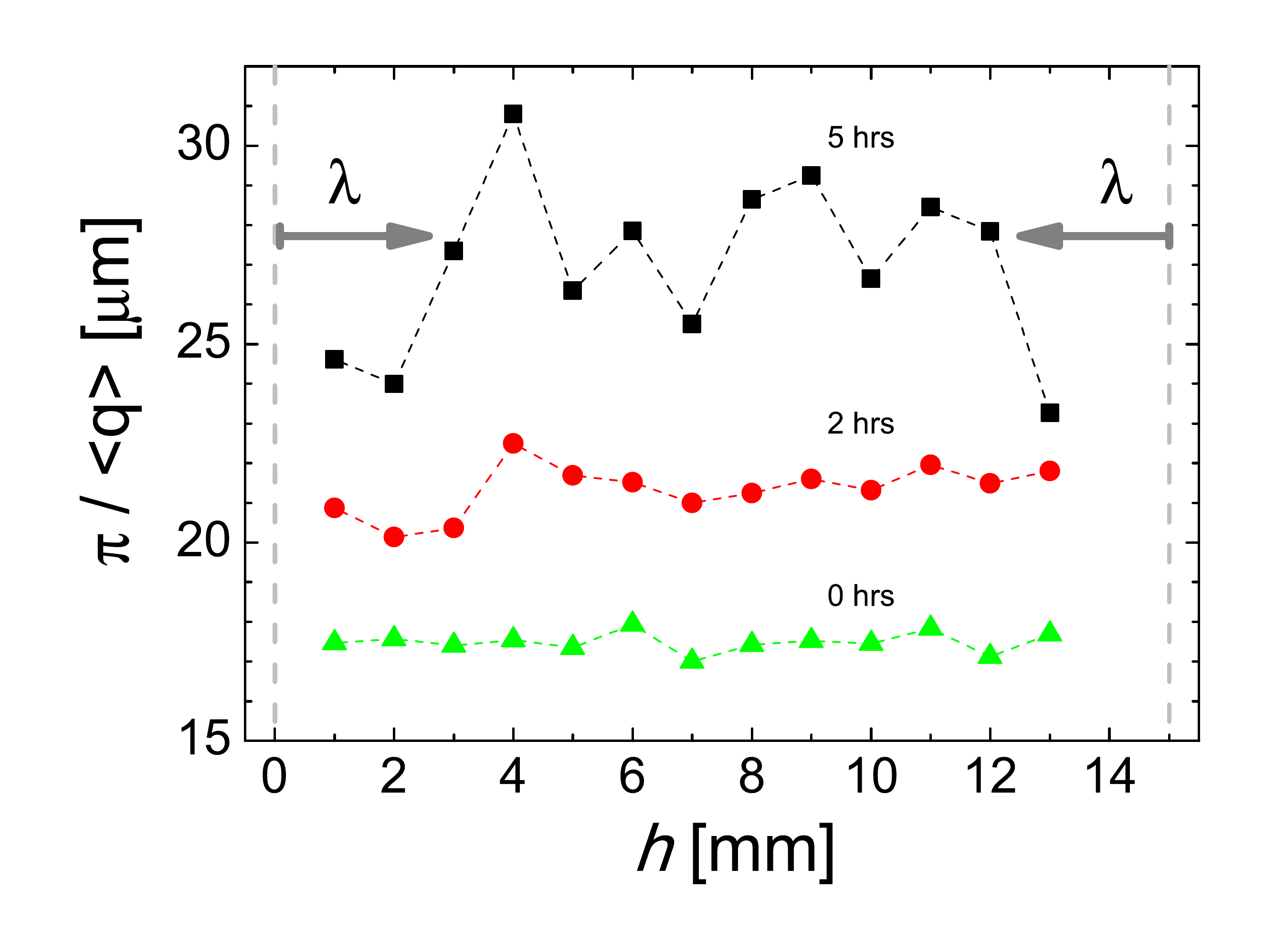}
        \caption{(Color online). The domain size $\R$ in a gel of height $\h0 = 15$ mm ($\cpstar = 2.4$) as a function of the distance measured from the base of the cell. Note
        aging is hindered at the boundaries of the gel. The effect continues over a distance $\lambda$ which is of order a few millimeters. }
        \label{fig:hetereogeneous}

\end{figure}

To test these ideas we have used confocal microscopy to monitor the time evolution of the network structure as a function of the vertical $z$-position within a gel. A series of 2D-confocal images were collected at regularly-spaced $1$ mm heights from a colloid-polymer gel with a total height of $\h0 = 15$ mm. Images were acquired for $\approx 7$ hours after the cessation of mixing, until the point at which gel collapse occurred. The characteristic domain size of the network $\R(h,\tw)$ at a height $h$ and time $\tw$  was calculated from the static structure factor $S(q,\tw)$
\begin{equation}\label{eqn:S(q)}
    S(q,\tw) = \frac{1}{2 \pi q \Delta q} \int_{q \leq |\mathbf{q'}| \leq q + \Delta q} \mathrm{d}\mathbf{q'} \left < \tilde{I}(\mathbf{q'},\tw) \tilde{I}(-\mathbf{q'},\tw) \right >
\end{equation}
where  $\tilde{I}(\mathbf{q},\tw)$ is the two-dimensional Fourier-transform of the image intensity $I(\mathbf{r},\tw)$ at time $\tw$, $\Delta q = 2 \pi /W$, and $W$ is the image width. The domain size is $\R = \pi /\q1$ where $\q1 = \int \textrm{d}q \; qS(q) / \int \textrm{d}q S(q)$. Measurements of $\R$ for different sample ages $\tw$ are plotted in Fig.~\ref{fig:hetereogeneous} and confirm that the aging of the gel network shows considerable spatial diversity: the domain size is large in some regions of space and small in others. Immediately after mixing, we observe the formation of a very uniform network with an average domain size of $\left < \R \right > = 17.5 \; \mu$m and a spatial variation of just 1.4\% (standard deviation/$\left < \R \right >$). But after $\tw = 2$ h, while the mean size has grown only slightly to $\left < \R \right > = 21.3 \; \mu$m the spatial variation in $\R$ has increased to 3\%. After 5 h, the variation in the domain size has increased still further to 9\% ($\left < \R \right > = 26.7 \; \mu$m). Clearly the data reveals that aging of the network is heterogeneous with spatial variation increasing with sample age.

\nomenclature[I]{$\tilde{I}$}{Fourier transform of image intensity}
\nomenclature[W]{$W$}{Image width}

\section{Discussion }
\label{sec:discussion}

The most striking feature of our results is the appearance during collapse of the $\frac{3}{2}$-power law dependence of the height $h(\ts)$ on the elapsed time $\ts$. A natural question is the physical origin of this unusual behavior. We propose that the dominant mechanism for collapse is the appearance of random micro-collapsed regions throughout the gel. On a microscopic level, the basic idea is that the particles comprising the gel attract each other relatively strongly so over time the gel spontaneously restructures locally to create small regions of more dense packing.  Since the collapsing particles are attached quite strongly to strands of the network as they rearrange they induce a local pressure field. This induced field, as a consequence of the poroelastic character of the gel, expands relatively slowly into the bulk of the gel. It is this long-range pressure field which we hypothesize generates the characteristic collapse dynamics evident in our experiments. Similar arguments have been invoked to account for the anomalous microscopic motion evident in dynamical light scattering of colloidal gels \cite{8739,4480} but not, as far as we are aware, for the macroscopic settling dynamics of gels. While our discussion has some features in common with the purely elastic models used previously \cite{8739,4480}, we focus here on deformations at large (macroscopic) length scales where poroelastic fluid flows are important.

For the moment, we idealize the gel as a one-dimensional chain of particles. Then if two particles leaves their equilibrium positions to stick together the left-hand particle imposes a force $+f_{0}$ on the left part of the chain at $z'$, while the right-hand particle imposes an equal and opposite force $-f_{0}$ on the right-hand side of the chain located at $z'+\delta z'$. The net effect of the rearrangement is therefore the creation of a local dipolar force $f(z',t')$ at the random position $z = z'$. The intensity of this dipolar force, the dipole moment $\mu$, is the product of the force $f_{0}$ and the displacement vector $\delta z'$ in the limit as $\delta z' \rightarrow 0$.   At the dipolar stress center the fluid pressure $P$ rises rapidly to a high value while further away $P$ is almost unchanged. The resulting pressure gradient drives a flow through the porous medium and as fluid exits from around the applied force the gel deforms locally and more of the load is borne by the network. Eventually, the pressure at all points reaches the same value and the gel relaxes so that the applied load is everywhere balanced by the elastic stresses in the network. The time scale for this equilibration is determined by a diffusion constant $\Ds$, with the deformation in the gel occurring over a length scale $ \approx \sqrt{\Ds t}$ in a time $t$. The gel diffusion constant $\Ds$ is \cite{157,9027}
\begin{equation}\label{eqn-d}
    \Ds = \frac{\elastic k}{\etaL (1-\phicolloid)}.
\end{equation}
where $k$ is the permeability of the network, $\etaL$ is the viscosity of the continuous phase, and $\elastic$ is the bulk modulus of the network. Using measurements of the low shear viscosity ($\etaL \simeq 0.1$ Pas), the elastic modulus of the gels ($\elastic \simeq 10$ Pa), and literature values for the permeability of similar density gels \cite{3708} ($k \sim 10 \a^{2} \simeq 10^{-12}$ m$^{2}$) we estimate a diffusion constant in our system of $ \Ds \simeq  10^{-10}$ m$^{2}$s$^{-1}$. Both the pore pressure and, in general, the permeability will change with time as the gel contracts locally. However in the initial stages of collapse the gel is uncompressed, the permeability is constant, and the equations of linear poroelasticity apply \cite{11914}. We shall ignore all non-linear effects. The bulk shrinkage of an unconstrained gel is linear so it is natural to assume that the dipole intensity will also be a linear function of time, $\mu(t) = \mu_{0}t$.

\begin{figure*}
        \includegraphics[width=0.9\textwidth,viewport=80 220 700 450]{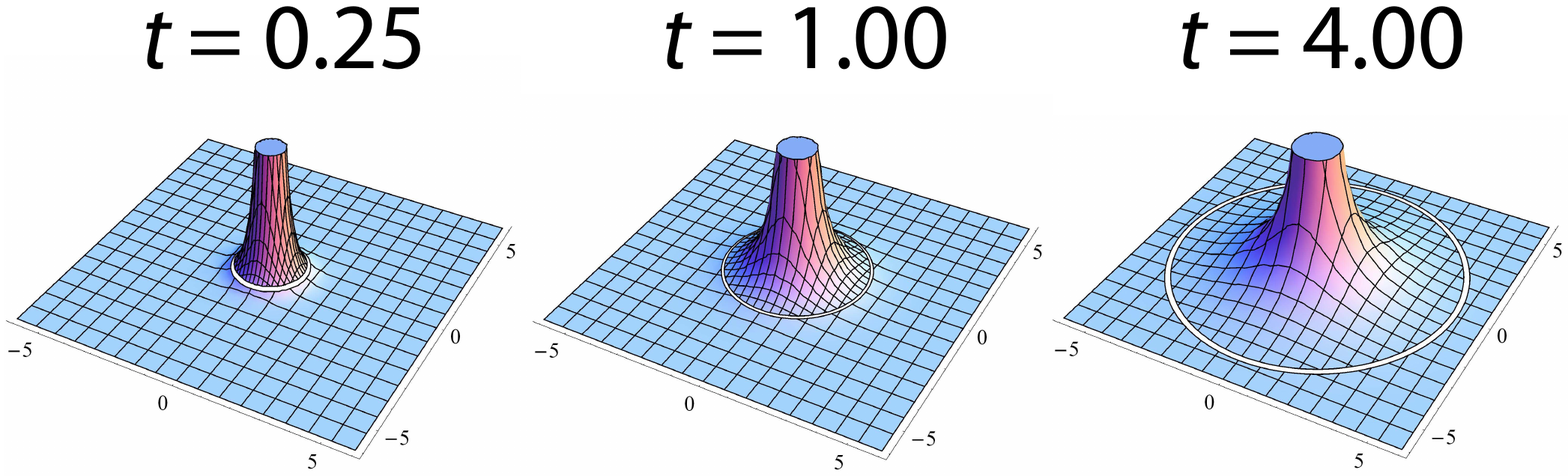}
        \caption{(Color online). An illustration of the pore pressure field generated by a continuous compression center placed at the origin of a gel. Times and distances are scaled so the diffusion constant $\Ds$ is unity. The ring of radius $r = 6 (\Ds t)^{1/2}$ parameterizes the spatial extent of the induced pressure field.}
        \label{fig:pressure}

\end{figure*}

For simplicity, we first consider the isotropic deformation produced in the gel by the supposition of three continuous orthogonal stress dipole centers (a single center of compression \cite{Love}). Rudnicki \cite{11949} has shown that the pore pressure at a distance $r$ from a single continuous center of compression in a fluid-saturated porous solid is of the form
\begin{equation}\label{eqn-P}
    P_{s}(r,t) = \frac{Q}{4 \pi \Ds r }  \textrm{erfc}(\xi/2),
\end{equation}
where $\xi = r / (\Ds t)^{1/2}$, $Q$ is proportional to $\mu_{0}$, and erfc is the complementary error function \cite{Abramo}. As $t \rightarrow 0$, $\textrm{ercf}(r / 2\sqrt{\Ds t}) \rightarrow 0$, and the pressure is zero. At finite times, the fluid has an opportunity to diffuse away from the origin, and the spherically symmetric pressure wave expands. In Fig.~\ref{fig:pressure} we plot a time series of the spreading pressure field as it diffuses away from the origin. The length scale where the pressure is finite is controlled by fluid diffusion within the gel and is thus time-dependent. From the figure it is clear that the spatial extent of the pressure field is of order $6 (\Ds t)^{1/2}$. Finally, as $t \rightarrow \infty$, $\textrm{ercf}(r / 2\sqrt{\Ds t}) \rightarrow 1$, and the pressure field approaches the pure $1/r$-elastostatic solution, expected for a continuous dipole source \cite{11914}.

We assume that at the start of collapse the gel contains centers of compression which are randomly distributed throughout the height of the gel, with a certain number $\rho$ per unit length. To calculate the velocity of the gel interface $\partial_{t} w$ we note that, from Darcy's law (Eq.~\ref{eqn-darcy}),  $\partial_{t} w$ is fixed by the total pressure gradient at the bottom of the gel. If we assume that each micro-collapse contributes independently then the pressure gradient is simply the sum of the gradients generated by individual events occurring at different heights $z_{j}$ within the gel. For simplicity we ignore the true vectorial nature of the problem and use a scalar approximation in which we assume that the pressure field due to each independent micro-collapse adds together coherently to produce the macroscopic pressure field. Since
the vertical pressure gradient must control the process of collapse we focus solely on the $z$-direction. A single center of compression located a distance $z_{j}$ from the gel interface creates a pressure gradient, $\theta_{j} = \partial_{r} P_{s}$, which from Eq.~\ref{eqn-P} is
\begin{eqnarray}\label{eqn-theta-j}
    -\theta_{j}(z_{j},t) & = &  \frac{Q}{4 \pi z_{j}^{2} \Ds  }  \textrm{erfc}(\frac{z_{j}}{2\sqrt{\Ds t}}) \nonumber \\
                         &  & + \frac{Q \textrm{e}^{- \frac{z_{j}^{2}}{4\Ds t}}}{4 \pi^{3/2} z_{j} \Ds^{3/2} \sqrt{t}  }.
\end{eqnarray}
The growth data in Fig.~\ref{fig:hetereogeneous} suggests that coarsening of the network is suppressed near a surface, \textit{i.e.} micro-collapse events appear preferentially at distances $z \geq \lambda$ away from a surface. This seems plausible since the energy barrier for a rearrangement near an interface will probably be higher than for the same event in the bulk of the gel, as the strain induced by the creation of the dipole is larger. Inspection of the data in Fig.~\ref{fig:hetereogeneous} suggests that $\lambda$ is of order a few millimeters. Hence, we assume that micro-collapses are uniformly distributed over the interval from $z = \lambda$ to $z = \h0-\lambda$. The total pressure gradient at the base of the gel is therefore,
\begin{equation}\label{eqn-pressure-grad}
 \partial_{z} P \Big |_{\textrm{base}} = \rho \int_{\lambda}^{\h0-\lambda} \theta(z,t) \textrm{d}z .
\end{equation}
Since experimentally we have observed that the collapse process does not vary with the total height $\h0$ of the gel the upper limit of the integral can be extended to $z = \infty$ without significant error. After inserting Eq.~\ref{eqn-theta-j}, the resulting integral can be performed exactly with the result:
\begin{eqnarray}\label{eqn-surface-pressure}
     -\partial_{z} P(t)  & = & \frac{\rho Q }{4 \lambda^{2} \pi^{3/2} \Ds^{1/2}  } \sqrt{t} \textrm{e}^{- \frac{\lambda^{2}}{4\Ds t}} \nonumber \\
                                              &   & + \frac{\rho Q \ln \lambda }{4 \pi^{3/2} \Ds^{3/2} \sqrt{t} } .
\end{eqnarray}
In the regime where $t \gg \lambda^{2}/ 4 \Ds$, which from our estimates for $\Ds$ and $\lambda$ equates to $t \gg 10^{3}$ s and holds for all but the shortest times studied, the expression for the pressure gradient simplifies to
\begin{equation}\label{eqn-asymptotic-pressure-gradient}
     -\partial_{z} P(t) \underset{t \gg \lambda^{2}/ 4 \Ds}{\approx} \;\; \frac{\rho Q }{4 \lambda^{2} \pi^{3/2} \Ds^{1/2}  } \sqrt{t}.
\end{equation}
On timescales $t \approx 10^{3}$ s the gel has typically not collapsed to any significant degree (see for instance the data in Fig.~\ref{fig:collapse-dynamics}) so
the permeability of the gel is not substantially changed from its initial value and $\Ds$ is time-independent.

The confocal data, presented in Sec.~\ref{sec:origin-delay}, reveals that micro-collapses first appear in gels with an age $\tw$ of $\approx \taud$ so the time $t$ available for the propagation of the pressure wave in Eq.~\ref{eqn-asymptotic-pressure-gradient} is $\tau = \tw - \taud$. By combining Eq.~\ref{eqn-darcy}, \ref{eqn-continuity}, and \ref{eqn-asymptotic-pressure-gradient} we obtain
\begin{equation}\label{eqn-theory-delta-h}
    \Delta h (\tau) = \left [\frac{\rho k Q }{4 \eta(1-\phicolloid) \lambda^{2} \pi^{3/2} \Ds^{1/2}  } \right ] \tau^{3/2}.
\end{equation}
where we have assumed that the flow of fluid through the network determines the initial rate of collapse, and $t \gg \lambda^{2}/ 4 \Ds$. This expression is in very good agreement with the experimental results where a similar exponent of 3/2 has been found, thus providing convincing evidence for our simple phenomenological model. In addition, our arguments predict that the coefficient of $\tau^{3/2}$, which we identify with the scale factor $A$ in Eq.~\ref{eqn-experimental-t3/2}, should be a system constant, independent of the initial height of the gel. This agrees with the height-independence seen in the experimental data presented in Fig.~\ref{fig:t3/2-plots}(c). We expect that the formation of micro-collapses is thermally activated so their number density $\rho$ will be of the form
\begin{equation}\label{eqn-number-density}
    \rho = \rho_{0} \; \mathrm{exp} \left (  - \frac{\Delta E}{\kBT} \right )
\end{equation}
where $\Delta E$ is an energy barrier for rearrangement. Since $\Delta E / \kBT $ will scale with the strength of the interparticle potential $\Uc$ one expects that the scale factor $A$ will depend exponentially on the interparticle potential, or equivalently the polymer concentration, in agreement with the experimental data plotted in Fig.~\ref{fig:t3/2-plots}(d). Finally, we note that our model displays no explicit dependence on gravity since we hypothesize that gel collapse is a consequence of irreversible aging of the particle network. Gravity simply dictates the direction of gel collapse and we believe that, in the current case, the gravitational stress on the network is not sufficiently large to significantly enhance the thermal relaxation of particle bonds. However, this conjecture still awaits a direct experimental proof. A systematic investigation of the collapse of transient gels as a function of the gravitational stress, which could be achieved by for example changing the density mismatch $\Delta \rho$ or by using micro-gravity conditions, would confirm this prediction. Unfortunately, no such data is currently available although we plan in the near future to start such measurements.

\nomenclature[\rho]{$\rho$}{Number of micro-collapses per unit length}%
\nomenclature[\theta]{$\theta$}{Pressure gradient due to single stress center}%
\nomenclature[\lambda]{$\lambda$}{Distance from free interface where stress centers excluded}%
\nomenclature[\Delta E]{$\Delta E$}{energy barrier for micro-collapse}%

\nomenclature[\xi]{$\xi$}{Scaled radial coordinate $= r / \sqrt{\Ds t}$}
\nomenclature[r]{$r$}{radial coordinate}
\nomenclature[Q]{$Q$}{Fluid volume injected}
\nomenclature[S]{$S$}{Dynamic structure factor}
\nomenclature[q]{$q$}{Wavevector}
\nomenclature[Ds]{$\Ds$}{Stress diffusion constant \verb+\Ds+}
\nomenclature[K]{$\elastic$}{Drained bulk modulus of gel}
\nomenclature[k]{$k$}{Permeability of gel}

\section{Summary}
\label{sec:summary}

We have studied the gravitational collapse of a colloidal gel by a combination of confocal microscopy, time-lapse video imaging, and rheology focusing particularly on the effect of the initial height $\h0$ of the gel and the strength of attractions $\Uc$.  The gels are made of emulsion drops suspended in a refractive index-matched mixture of ethylene glycol and water, with a high molecular polymer added to induce a weak long-range attraction. For all systems, the height $h(\tw)$ of the gel shows a characteristic two-step decay as a function of age $\tw$: for $\tw$ less than the lag time $\taud$ the system resists gravity and there is no significant deformation, but for $\tw > \taud$ the gel abruptly yields and collapses. The change in the height $\Delta h = \h0 - h(\tw)$ of the gel during collapse has a number of distinctive features. First, we find that the initial degree of settling is well described by the expression, $\Delta h \sim \ts^{3/2}$, with $\ts$ the time counted from the moment when collapse first starts. Second, both the process of collapse and the lag time $\taud$ are independent of the initial height of the gel. Microscopically, the gel consists of a network of interconnected strands of particles. Confocal microscopy reveals that continuous restructuring of this network occurs which, with increasing age, leads to the breaking of bonds between particle strands and a progressive weakening of the network. The subsequent reduction in the large scale connectivity of the network eventually triggers a macroscopic collapse. Measurement of the microscopic structure of the gel during settling show that the age-dependent changes in the network are not isotropic but are concentrated around large inhomogeneities within the sample. We hypothesize that the collapse of the gel is determined by the rate at which fluid can be expelled from the gel. A simple phenomenological model of fluid flow driven by the formation of random compression centers within the gel correctly accounts for the behavior experimentally observed.

\begin{acknowledgments}

We thank C.P. Royall, R. Buscall, W. Poon, and W. Kob  for helpful discussions, Leila Wannell and Humphrey Yeung for their assistance with the experiments, and an anonymous referee for their insightful comments and suggestions. The work was supported by Bayer CropScience and the UK Engineering and Physical Science Research Council through the award of an Industrial Case Studentship from Chemistry Innovations KTN to LJT.

\end{acknowledgments}


\newpage

\nomenclature[rg]{$\rg$}{polymer radius of gyration \verb+\rg+}%
\nomenclature[a]{$\a$}{PDMS radius \verb+\a+}%
\nomenclature[cp]{$\cp$}{polymer concentration \verb+\cp+}%
\nomenclature[cpstar]{$\cpstar$}{scaled polymer concentration \verb+\cpstar+}%
\nomenclature[\phi0]{$\phic$}{Initial colloid volume fraction \verb+\phic+}%
\nomenclature[\phig]{$\phig$}{Colloid volume fraction in gel phase \verb+\phig+}%
\nomenclature[\phic]{$\phicolloid$}{colloid volume fraction \verb+\phicolloid+}%
\nomenclature[qr]{$\qR$}{polymer-colloid size ratio \verb+\qR+}%
\nomenclature[Uc]{$\Uc$}{potential at contact \verb+\Uc+}%
\nomenclature[U]{$\U$}{interparticle potential \verb+\U+}%
\nomenclature[\Delta]{$\D$}{range of potential \verb+\D+}%
\nomenclature[Rc]{$\R$}{Domain radius of spinodal pattern \verb+\R+}
\nomenclature[qmean]{$\q1$}{Characteristic wavevector \verb+\q1+}

%
%
\nomenclature[h0]{$\h0$}{initial height of gel \verb+\h0+}%
\nomenclature[d]{$\d0$}{cell diameter \verb+\d0+}%
\nomenclature[hf]{$\hf$}{final equilibrium height of gel \verb+\hf+}%
\nomenclature[\tauc]{$\tauc$}{collapse time \verb+\tauc+}%
\nomenclature[\taud]{$\taud$}{delay time \verb+\taud+}%
\nomenclature[tw]{$\tw$}{waiting time \verb+\tw+}%
\nomenclature[\tau]{$\ts$}{$\tw - \taud$ \verb+\ts+}%
\nomenclature[\nu]{$\vc$}{collapse velocity = dh/dt \verb+\vc+}%
\nomenclature[\beta]{$\betac$}{The stretching exponent \verb+\betac+}%
\nomenclature[\sigma]{$\sig$}{Applied stress on gel \verb+\sig+}
\nomenclature[\sigmag]{$\sigg$}{gravitational stress on gel \verb+\sigg+}
\nomenclature[\sigmay]{$\sigy$}{yield stress of gel \verb+\sigy+}
\nomenclature[nbreak]{$\nbreak$}{Number of break events \verb+\nbreak+}
\nomenclature[nlink]{$\nlink$}{Number of link events \verb+\nlink+}
\nomenclature[P]{$P$}{Fluid pressure}
\nomenclature[v]{$v$}{velocity of fluid flow through gel}
\nomenclature[w]{$w$}{Displacement of solid network}


\nomenclature[kb]{$\kb$}{Boltzmann constant \verb+\kb+}%
\nomenclature[kbt]{$\kBT$}{Thermal energy \verb+\kBT+}%


\end{document}